\documentclass[journal=jacsat,manuscript=article]{achemso}
\usepackage{graphicx}
\usepackage{amsmath,units}
\usepackage{soul}
\usepackage{babel}
\usepackage{mathrsfs}
\usepackage{xcolor} 
\usepackage[export]{adjustbox}
\usepackage{epstopdf}
\allowdisplaybreaks[1]

\def\bea{\begin{eqnarray}}
\def\eea{\end{eqnarray}}

\author{Ofir Tal-Friedman}
\affiliation{School of Physics \& Astronomy, Tel Aviv University, Tel Aviv 6997801, Israel}

\author{Tommer D. Keidar}
\affiliation{School of Chemistry, Tel Aviv University, Tel Aviv 6997801, Israel}
\alsoaffiliation{The Center for Computational Molecular and Materials Science, Tel Aviv University, Tel Aviv 6997801, Israel}
\alsoaffiliation{The Center for Physics and Chemistry of Living Systems, Tel Aviv University, Tel Aviv 6997801, Israel}

\author{Shlomi Reuveni}
\email{shlomire@tauex.tau.ac.il}
\affiliation{School of Chemistry, Tel Aviv University, Tel Aviv 6997801, Israel}
\alsoaffiliation{The Center for Computational Molecular and Materials Science, Tel Aviv University, Tel Aviv 6997801, Israel}
\alsoaffiliation{The Center for Physics and Chemistry of Living Systems, Tel Aviv University, Tel Aviv 6997801, Israel}

\author{Yael Roichman}
\email{roichman@tauex.tau.ac.il}

\affiliation{School of Chemistry, Tel Aviv University, Tel Aviv 6997801, Israel}
\alsoaffiliation{School of Physics \& Astronomy, Tel Aviv University, Tel Aviv 6997801, Israel}

\alsoaffiliation{The Center for Physics and Chemistry of Living Systems, Tel Aviv University, Tel Aviv 6997801, Israel}

\title[An \textsf{achemso} demo]
{Smart Resetting: An Energy-Efficient Strategy for Stochastic Search Processes}

\begin{document}
\begin{abstract}
\noindent
Stochastic resetting, a method for accelerating target search in random processes, often incurs temporal and energetic costs. For a diffusing particle, a lower bound exists for the energetic cost of reaching the target, which is attained at low resetting rates and equals the direct linear transportation cost against fluid drag. Here, we study ``smart resetting," a strategy that aims to beat this lower bound. By strategically resetting the particle only when this benefits its progress toward the target, smart resetting leverages information to minimize energy consumption.  We analytically calculate the energetic cost per mean first passage time and show that smart resetting consistently reduces the energetic cost compared to regular resetting. Surprisingly,  smart resting achieves the minimum energy cost previously established for regular resetting, irrespective of the resetting rate. Yet, it fails to reduce this cost further. We extend our findings in two ways: first, by examining nonlinear energetic cost functions, and second, by considering smart resetting of drift-diffusion processes.

\end{abstract}
\section{Introduction}
\noindent

A hallmark of resetting is that it can expedite stochastic search processes \cite{evans_diffusion_2011,pal_first_2017,chechkin_random_2018,rotbart2015michaelis,kusmierz_first_2014,bhat_stochastic_2016,yin2023restart}. For instance, introducing resetting to a diffusing particle prevents its mean first-passage time (MFPT) from diverging and renders it finite. Yet, this great advantage comes at a cost.
In most systems, resetting cannot be performed instantaneously; it takes time \cite{rotbart2015michaelis, Reuveni_2014_PNAS, pal2020search, bodrova_resetting_2020, Sunil_2023} and also requires energy \cite{Sunil_2023, Mori_2023_Entropy,tal-friedman_experimental_2020,fuchs2016stochastic}.
For example, resetting the position of particle diffusing in a viscous fluid requires working against the drag force. It can then be shown that the mean energetic cost per first passage event does not vanish at low resetting rates. Instead, it converges to a finite lower bound whose numerical value identifies with the energetic cost of simply dragging the particle from the origin directly to the target \cite{tal-friedman_experimental_2020,Sunil_2023}. As a result, even the introduction of an infinitesimally small resetting rate requires a jump in energy expenditure, indicating a qualitative departure from the behavior observed in the absence of resetting. The thermodynamic origin of this jump is still unclear.  

\begin{figure}[t]
	\centering
	\includegraphics[trim={3.9cm 3cm 4.5cm 1.7cm},clip,width=1\textwidth]{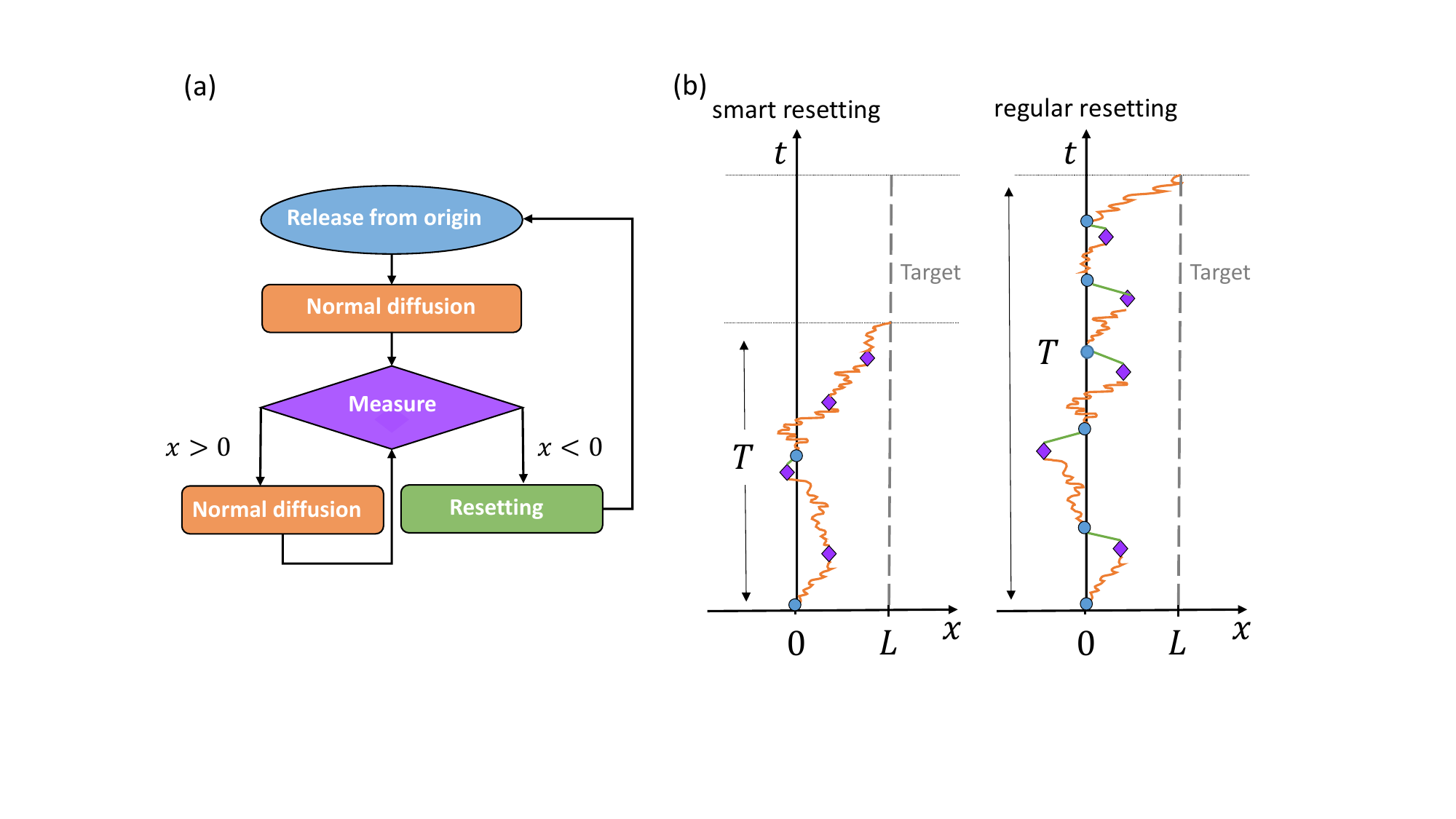}
	\caption{Diffusion with smart resetting. a) A particle is released from $x=0$ and allowed to diffuse towards a target at $L>0$. The particle's position is measured at a constant rate $r$, i.e., at exponentially distributed time intervals with mean $1/r$. After each measurement, the particle is reset back to $0$ only if its current location obeys $x(t)<0$, as described in the flow chart. b) Comparison between smart and regular resetting. In smart resetting (left panel), the particle is reset according to the scheme in panel (a), i.e., only if resetting brings the particle closer to the target. In regular resetting, the particle is reset regardless of its instantaneous position. }
	\label{fig:compare_trajectories}
\end{figure}

Inspired by results in information thermodynamics, we raise the following question: can information be used in order to reduce or even eliminate the energetic cost of resetting when this is used to facilitate search? For example, consider the ``smart resetting" model illustrated in Fig.~\ref{fig:compare_trajectories}. In contrast to regular resetting, in which the particle is reset with a constant rate $r$, smart resetting measures the particle's position at rate $r$ and only resets if this brings the particle closer to the target. This information-based feedback into the resetting process is expected to reduce both the MFPT and the energetic cost of reaching the target. However, the precise extent of this improvement remains to be addressed.   

Here, we study the temporal and energetic cost of smart resetting in the paradigmatic model of a particle diffusing in a viscous fluid. We find that smart resetting dramatically affects the energetic cost of finding a target; it is reduced and becomes resetting rate independent. Moreover, it always attains the lower bound on the energetic cost of regular resetting. 

We begin by introducing the model of smart resetting and derive a closed-form solution for the Laplace transform of the time-dependent distribution of this process in the presence of an absorbing target. We then use this expression to find the FPT distribution of the process and rederive the known MFPT. Having calculated the propagator of the smart resetting process, we derive an expression for its energetic cost per first passage and compare it to the cost of regular resetting. Finally, we analyze the case of drift-diffusion under smart resetting and use our newly found insights in order to explain its energetic cost per first passage.

\section{Search under smart resetting}
\noindent We start by analyzing the smart resetting model depicted in Fig.~\ref{fig:compare_trajectories}.This model belongs to a larger class of resetting processes with position-dependent resetting rates. These were studied in different contexts previously, including biophysics \cite{tucci2020, Mori2023, Bressloff_2023}, non-equilibrium steady states \cite{Falcao_2017, roldan2017path}, control theory \cite{DeBruyne2023, de_bruyne_optimization_2020}, catastrophic events \cite{catastrophic_resetting, Ali_2022}, as a shortcut to confinement \cite{roldan2017path}, and as means to accelerate the sampling of long timescales in MD simulations \cite{church2024}.

We will first assume that resetting results in an instantaneous return to the origin. Later on, we will show that the temporal and energetic cost of non-instantaneous returns can be obtained from the solution to the instantaneous model. Smart resetting with instantaneous returns to the origin can be described by the following master equation for the probability distribution of the position of the particle,
\begin{equation}\label{eqn: fokker planck}
    \frac{\partial p(x,t)}{\partial t}=D\frac{\partial^2 p(x,t)}{\partial x^2}-rp(x,t)\Theta(-x)+r\delta(x)Q(t).
\end{equation}

\noindent Here, $D$ is the diffusion coefficient, $r$ is the resetting rate, $\Theta(x)$ is the Heaviside function, and $Q(t)\equiv\int_{-\infty}^0 p(x,t) \, dx$. The second term on the right-hand side is associated with probability loss due to resetting. The Heaviside function encodes the spatial dependence of the conditioned resetting, allowing only resetting events originating at $x<0$. The third term on the right-hand side is the probability gain at the origin. 

We solve Eq.~\eqref{eqn: fokker planck} with an initial condition of  $p(x,0)=\delta(x)$, and an absorbing boundary at $x=L$. Laplace transforming Eq.~\eqref{eqn: fokker planck} we obtain
\begin{equation}\label{fokker planck laplace}
    \frac{\partial^2\Tilde{p}(x,s)}{\partial x^2}=\frac{r\Theta(-x)+s}{D}\Tilde{p}(x,s)-\frac{r\Tilde{Q}(s)+1}{D}\delta(x).
\end{equation}

We then solve this differential equation and find (See SI)
\begin{equation}\label{eqn:propagator_in_laplace_space}
\begin{split}
    &\Tilde{p}(x,s)=\\
    &\begin{cases}
        \frac{e^{\sqrt{\frac{s}{D}}(2L-x)}-e^{\sqrt{\frac{s}{D}}x}}{\sqrt{sD}\left[e^{2\sqrt{\frac{s}{D}}L}\left(1+\sqrt{\frac{s}{s+r}}\right)+1-\sqrt{\frac{s}{s+r}}\right]}& \text{if } 0<x<L\\
        \frac{\left(e^{2\sqrt{\frac{s}{D}}L}-1\right)e^{\sqrt{\frac{s+r}{D}}x}}{\sqrt{sD}\left[e^{2\sqrt{\frac{s}{D}}L}\left(1+\sqrt{\frac{s}{s+r}}\right)+1-\sqrt{\frac{s}{s+r}}\right]} &\text{if } x<0.
    \end{cases}
\end{split}
\end{equation}

\begin{figure}[t!]
	\centering
	\includegraphics[trim={8.9cm 0 11.8cm 0},clip,width=0.5\textwidth]{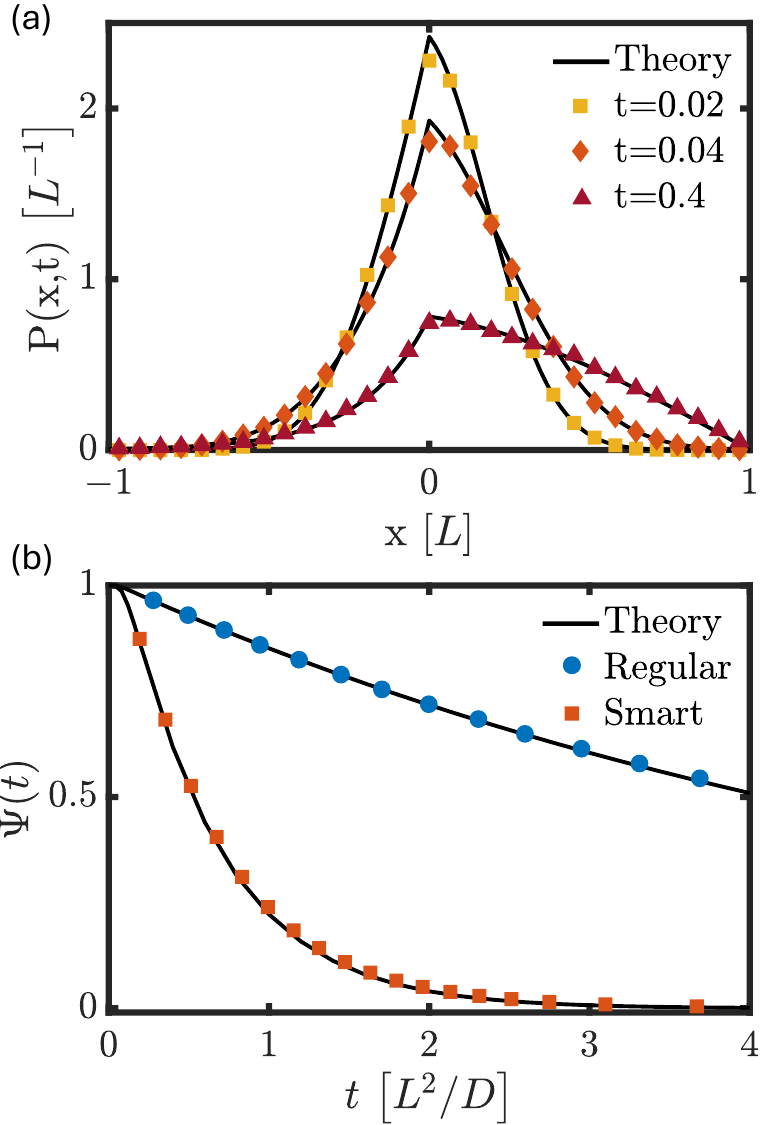}
	\caption{(a) Time-dependent probability distribution of diffusion with smart resetting. Results are shown for three different times (measured in units of $L^2/D$). Very good agreement is seen between simulations (symbols) and theory (solid lines - the numerical inverse of Eq.~\eqref{eqn:propagator_in_laplace_space}).  (b) The survival function of diffusion with regular and smart resetting. Symbols represent simulation results, and solid lines represent numerical Laplace inversion of theory; for regular resetting based on Evans \& Majumdar \cite{evans_diffusion_2011}, and for smart resetting based on Eq. (\ref{eqn:survival_laplace}).
    Here, the resetting rate is $r=25 \left[ \frac{D}{L^2} \right]$.}
	\label{fig:time_dependent_propagator_fig}
\end{figure}

To corroborate  Eq.~\eqref{eqn:propagator_in_laplace_space}, we numerically perform the inverse Laplace transform to obtain the propagator in real time. In parallel, we perform Langevin dynamics simulations of a diffusive searcher with smart resetting to obtain directly the real-time propagator. Here and throughout the paper, we took $D=1$ and set the absorbing boundary at $x=L=50$. We find perfect agreement between the analytical and numerical results (Fig.~\ref{fig:time_dependent_propagator_fig}{a}). As expected, $p(x,t)$ is asymmetric with respect to the origin, and its height diminishes with time. To quantify absorbance at the target, we also compute the survival probability $\Psi (t)=\int_{-\infty}^L p(x,t)dx$. This is readily done in Laplace time by substituting Eq.~\eqref{eqn:propagator_in_laplace_space} and performing the integral. We obtain  
\begin{equation}
\begin{split}
  \Tilde{\Psi} (s) =&
   \frac{1}{s+\sqrt{s(s+r)}\coth{\gamma}}\\ +&\frac{\cosh{\gamma}-1}{s\left(\cosh{\gamma}+\frac{s}{\sqrt{s(s+r)}}\sinh{\gamma}\right)},
   \end{split}
   \label{eqn:survival_laplace}
\end{equation}

\noindent where $\gamma\equiv\sqrt{\frac{s}{D}}L$ (see SI). We then compare the survival probability for regular resetting and smart resetting (Fig.~\ref{fig:time_dependent_propagator_fig}b). The results align with our intuition: particles find the target much faster under smart resetting. Moreover, the speedup can be significant at high resetting rates. For example, in Fig.~\ref{fig:time_dependent_propagator_fig}b, at time $t=4 \,
[L^2/D]$, only about half of the particles undergoing regular resetting have arrived at the target. In contrast, almost all particles undergoing smart resetting have already arrived. 

\begin{figure}[t]
	\centering
	\includegraphics[trim={8.9cm 0 10cm 0},clip,width=0.5\textwidth]{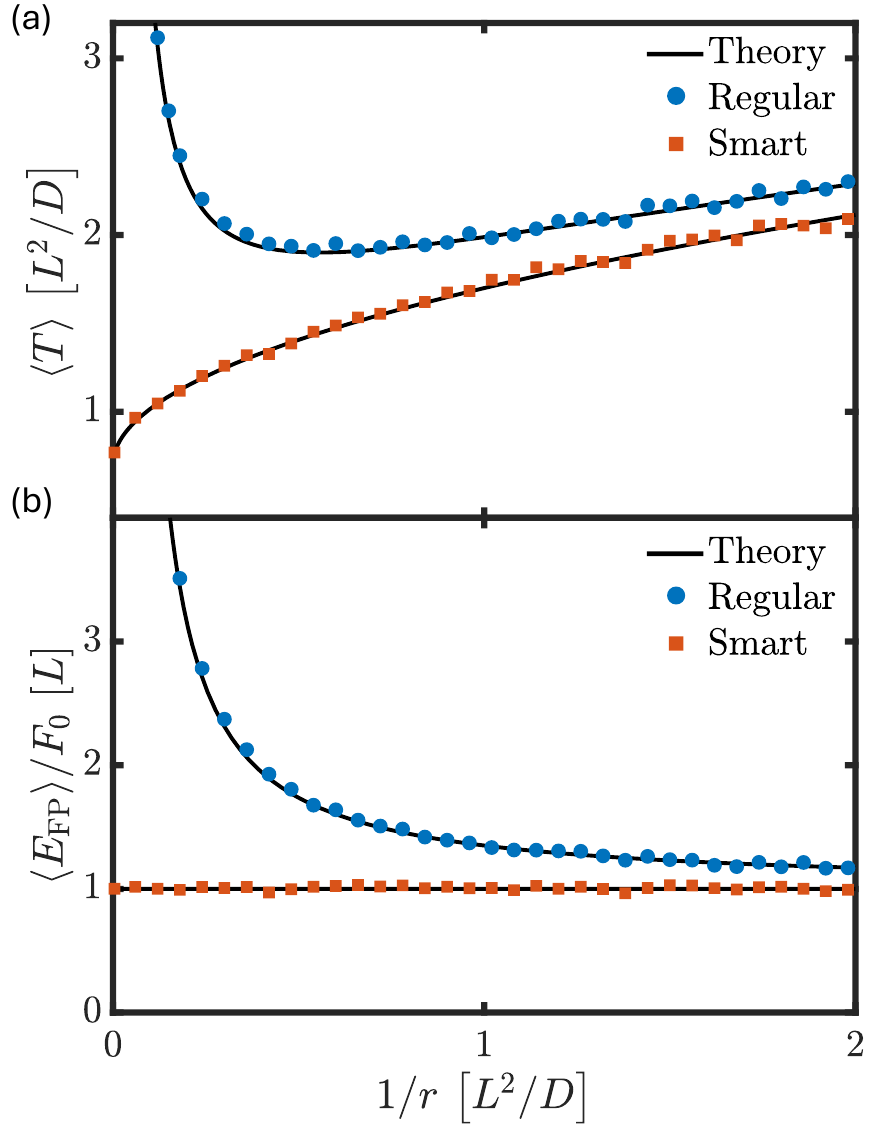}
	\caption{(a) A comparison of the MFPT as a function of $1/r$ of regular (blue circles) and smart (orange squares) resetting, with constant velocity returns.  Symbols come from Langevin simulations. The lines come from Eq.~\ref{eqn:MFPT_smart_constant_velocity} (smart resetting) and Eq. 9 in \cite{pal2020search} (regular resetting).  (b) The average energetic cost per first passage as a function of $1/r$ for regular (blue circles) and smart (orange squares) resetting. The lines represent theory, and the symbols come from Langevin simulations. For smart resetting, it can be seen that for every value of $1/r$ the energetic cost is exactly $F_0L$. Here, the return speed is set to $5 \left[ \frac{D}{L} \right]$.}
 \label{fig:MFPTandWork}
 \end{figure}

This reasoning holds true for all resetting rates, as can be ascertained from the dependence of the MFPT on the resetting rate. Under smart resetting, the MFPT is given by $\langle T\rangle = \frac{L^2}{2D}+\frac{L}{\sqrt{Dr}}$\cite{catastrophic_resetting} (also see SI),  which is always smaller than the MFPT for regular resetting \cite{evans_diffusion_2011}, $\langle T\rangle = (e^{\sqrt{\frac{r}{D}}L}-1)/r$. 
In the limit of $r\ll \frac{D}{L^2}$, very few measurements occur, and most of them `catch' the particle at $x<0$, i.e., a measurement will almost surely result in resetting. It is thus no wonder that in this limit the MFPTs of smart and regular resetting are asymptotically equal. In the other limit, high resetting rates, the MFPT of regular resetting diverges while the MFPT under smart resetting converges to a constant value of $L^2/2D$. The latter is precisely the MFPT for a diffusing particle starting at the origin and diffusing to a target at a distance $L$ with a reflecting boundary at the origin.
Similar results are obtained when the temporal cost is taken into account (Fig.~\ref{fig:MFPTandWork}a); the MFPTs of smart and regular resetting are asymptotically equal in the limit of small resetting rates and the MFPT under smart resetting converges to a constant.

\section{Energetic and temporal cost} 
\noindent
We now compute the energetic cost of smart resetting. Intuitively, we expect that the elimination of counterproductive returns in smart resetting will translate to a reduction in the energetic cost per first passage event. To show this, we consider systems in which the energetic cost of returns, $E=F_0|\Delta x|$, is linear in the drag distance, $|\Delta x|$, with a proportionality constant $F_0$ that depends on system parameters, e.g., colloidal particles returned to the origin at a constant speed against a friction force of a viscous fluid. In such systems the mean energetic cost per arrival at the target for regular stochastic resetting is \cite{tal-friedman_experimental_2020}
\begin{equation}
    \label{eqn:energy_first_passage_resetting_norm}
    \langle E_\text{FP} \rangle = F_0 L\left[\frac{2\sinh(\alpha_0 L)}{\alpha_0 L}-1\right],
\end{equation}
with  $\alpha_0=\sqrt{r/D}$. Observe that as $r$ decreases, the mean energetic cost decays toward $F_0L$ (Fig.~\ref{fig:MFPTandWork}b). Remarkably, this lower bound is the energetic cost of bringing the searcher from the origin straight to the target \cite{tal-friedman_experimental_2020}.

The energetic cost per arrival at the target under smart resetting is calculated in the SI. Here, we only review the main steps. First, we define the rate of energy consumption, 
\begin{equation}\label{lambda1}
    \lambda(t;r)=\int_{-\infty}^0 -rF_0 xp(x,t;r) dx,
\end{equation}
where $-rF_0 x$ is the cost per unit time of returning the particle from position $x$ back to the origin, which is integrated over all $x<0$. Next, we integrate over time to obtain the mean energetic cost per first-passage event,
\begin{equation}\label{tot energy}
\langle E_\text{FP} \rangle=\int_0^\infty\lambda(t;r)\, dt.
\end{equation}
Representing this integral as the value of a Laplace transform in the limit $s\rightarrow0$ and using Eq.~\eqref{lambda1}, we obtain
\begin{equation}\label{Tot energy laplacec}
    \langle E_\text{FP} \rangle=-rF_0\int_{-\infty}^0 x\Tilde{p}(x,s\to0;r)\, dx.
\end{equation}
Plugging in the Laplace time propagator in Eq.~\eqref{eqn:propagator_in_laplace_space}, we obtain the mean energetic cost (per first-passage event) for diffusion with smart resetting
\begin{equation}
    \label{eqn:mean_energy_smart}
    \langle E_\text{FP} \rangle = F_0 L.
\end{equation}
As expected, the mean energetic cost per arrival at the target is always smaller than that under regular resetting (Fig.~\ref{fig:MFPTandWork}b). Surprisingly, however, it is independent of the resetting rate and always equal to the lower bound on the energetic cost of regular resetting.

In smart resetting, we leverage information to avoid counterproductive work by eliminating returns that drag the particle away from the target. A higher resetting rate implies more measurements and information utilization. While information and feedback enable reaching the target with minimal energy expenditure regardless of the resetting rate, a fundamental minimal cost is still required. This is a key result of our work.

Explicitly, we can analyze the difference in energetic cost between regular and smart resetting by looking separately at the contribution of returns toward the target $\langle \ell_t \rangle=\langle \sum_{\Delta x >0}|\Delta x| \rangle$ and away from it $\langle \ell_a \rangle=\langle \sum_{\Delta x <0}|\Delta x| \rangle$. The average energetic cost is $\langle E_\text{FP} \rangle=F_0 (\langle \ell_t \rangle+\langle \ell_a \rangle)$. For regular resetting we find (see SI),

\begin{equation}
    \label{eqn:cost toward and away}
\begin{split}
\langle \ell_t\rangle &= \sqrt{\frac{D}{r}}\sinh{\left(\sqrt{\frac{r}{D}}L\right)}, \\
\langle \ell_a\rangle &= \sqrt{\frac{D}{r}}\sinh{\left(\sqrt{\frac{r}{D}}L\right)} -L.
\end{split}
\end{equation}
Hence,
\begin{equation} \label{eqn: w_t-w_a=L}
    \langle \ell_t\rangle-\langle \ell_a\rangle=L.
\end{equation}
Namely, the difference between productive and counterproductive drags \emph{toward the origin} equals exactly the distance to the target.
Notice that this relation also holds for the case of smart resetting, where for any resetting rate, the total average dragging distance is $L$, and by definition, only productive drags are performed; hence, $\langle \ell_t\rangle = L$ and $\langle \ell_a\rangle = 0$ and $\langle E_\text{FP} \rangle=F_0L$ (Eq. \eqref{eqn:mean_energy_smart}).

Since in smart resetting, the mean dragging distance per first passage event is $\langle \ell\rangle = L$, the average duration spent in the return phase is $L /v_{r}$ where $v_{r}$ is the return speed. Adding this constant duration to the MFPT of smart resetting with instantaneous returns we obtain the MFPT of constant speed returns, 
\begin{equation}
\label{eqn:MFPT_smart_constant_velocity}
    \langle T\rangle = \frac{L^2}{2D}+\frac{L}{\sqrt{Dr}} + \frac{L}{v_r}.
\end{equation}
Note that the temporal cost added due to returns does not change the fact the optimal resetting is obtained at the limit of $r \to \infty$. This is in contrast to regular resetting, where the temporal cost due to constant speed returns decreases the optimal resetting rate.  

So far, we have only considered a cost that is linear in the drag distance. For a general power-law cost, in which the energy required to drag the particle a distance $\ell_0$ is $\kappa \ell_0^{\alpha}$, we find (see SI) 
\begin{equation}
    \label{eqn:mean_energy_smart_power_law}
\langle E_\text{FP}\rangle=\kappa L\Gamma(\alpha+1)\left(\frac{D}{r}\right)^{\frac{\alpha-1}{2}}.
\end{equation}
Interestingly, we find that the  cost of smart resetting at an arbitrary resetting rate is identical to the cost of regular resetting in the limit $r \ll 1$. Indeed,  expanding the latter to leading order in the resetting rate we obtain Eq. \eqref{eqn:mean_energy_smart_power_law} (see SI). Thus, while the energetic cost of smart resetting is no longer constant in the resetting rate, it can still be seen as a truncated version of the series expansion of the cost of regular resetting, as we have found for linear costs. Note that a similar result follows immediately for temporal costs where the time required to drag the particle a distance $\ell_0$ is $\kappa \ell_0^{\alpha}$, with $\kappa$ now taken to have units of time over length to the power alpha.


\section{Smart resetting of a drift-diffusion process}
\begin{figure}[t]
	\centering
	\includegraphics[trim={9.13cm 0 9.9cm 0},clip,width=0.5\textwidth]{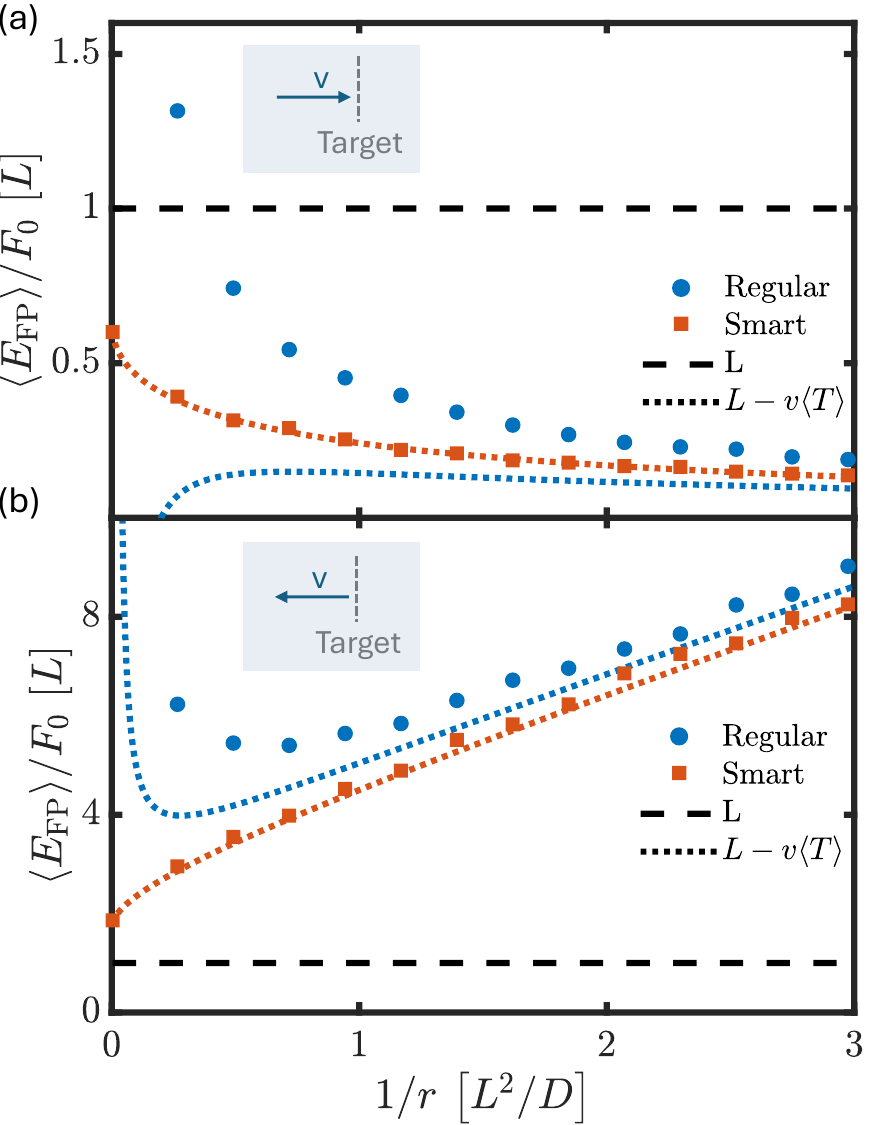}
	\caption{The average energetic cost per first passage for drift diffusion with $Pe=0.5$ (panel a) and $Pe=-0.5$ (panel b), as a function of $1/r$ for regular (blue) and smart (orange) resetting. The lines represent $L-v\langle T \rangle$, the symbols are Langevin simulations. As is expected from Eq. \eqref{eqn: w_t-w_a=L drift}, for smart resetting the normalized average energetic cost is equal to $L-v\langle T \rangle$, while for regular resetting, the cost exceeds this lower bound.}
 \label{fig:drift_drag}
 \end{figure}

\noindent
To better understand the lower bound of the energetic cost per first passage time, we study the case of drift-diffusion under regular stochastic resetting. Here, when the drift is directed toward the target, the MFPT is finite, even without resetting. Thus, in contrast to the no-drift case, there is no minimal bound on the energy cost required to ensure a finite MFPT. Nonetheless, regular resetting expedites first passage when the process is diffusion-dominated, i.e., for Péclet numbers $Pe\equiv Lv/2D$ between 0 and 1, where $v$ is the drift velocity \cite{ray_peclet_2019}. This comes at a cost, which depends on the resetting rate (Fig.~\ref{fig:drift_drag}a - blue circles). In contrast, for the case of $Pe<0$ (drift away from the target), resetting is needed to ensure arrival at the target and a finite MFPT. The required energy expenditure is then non-monotonic in the resetting rate (Fig.~\ref{fig:drift_drag}b - blue circles). 

Smart resetting is more energetically efficient than regular resetting, regardless of the drift direction and the resetting rate (Fig.~\ref{fig:drift_drag} - orange squares). This is expected intuitively, since it prevents non-beneficial returns. However, unlike our result for normal diffusion processes, the energetic cost is not constant, and depends on the resetting rate. 

One way to generalize Eq. \eqref{eqn: w_t-w_a=L} is to subtract the contribution of the drift from $L$ in the right-hand side of the equation,
 \begin{equation} \label{eqn: w_t-w_a=L drift}
    \langle \ell_t\rangle-\langle \ell_a\rangle=L-v\langle T \rangle.
\end{equation}
We corroborate Eq. \eqref{eqn: w_t-w_a=L drift} both for regular and for smart resetting in the SI.
As was previously discussed, the normalized mean energetic cost is simply $\langle \ell_t \rangle+\langle \ell_a \rangle$. Combining with Eq. \eqref{eqn: w_t-w_a=L drift} we obtain $\langle E_\text{FP} \rangle/F_0 =L-v\langle T \rangle+ 2\langle \ell_a \rangle$. We note that $\langle \ell_a \rangle \geq 0$, thus $L-v\langle T \rangle$ bounds the energy cost from below. This lower bound is achieved when $\langle \ell_a \rangle = 0$, such as in smart resetting, Fig. 4 (dashed lines). The MFPT used in Eq. \eqref{eqn: w_t-w_a=L drift} depends on the resetting protocol. For regular resetting, it is given in \cite{ray_peclet_2019}, and for smart resetting, it is derived in the SI. 

From a thermodynamic perspective, the drift force performs work on the searching particle, moving it closer or farther from the target. Consequently, the work invested to reach the target through resetting changes accordingly. When the drift is toward the target and motion is diffusion-dominated, the reduction in the MFPT achieved by resetting incurs an additional energy expenditure proportional to the MFPT and drift velocity. Conversely, if the drift points away from the target, the resetting energetic cost increases to compensate for the work performed by the drift force in displacing the particle from the target.  Even though the resetting rate does not appear explicitly in Eq.~\ref{eqn: w_t-w_a=L drift}, its effect on the dragging distance and hence on the energetic cost is manifested through the MFPT.

\section{Conclusion}
\noindent
In this paper, we have studied how information and feedback can further expedite a diffusive search process under resetting. Naturally, eliminating the counterproductive resetting events by smart resetting led to a reduction of the MFPT and the energetic cost of reaching the target. More striking is the fact that this energetic cost, when linear in drag distance, is independent of the resetting rate and coincides with the lower bound found previously for regular resetting \cite{tal-friedman_experimental_2020}. The latter is simply the cost of dragging the particle directly to the target.  Remarkably, the formulas for the temporal and the energetic costs of smart resetting emerge as truncated versions of the series expansions derived for regular resetting, which is also true for power-law costs. 

Generalizing our analysis to drift-diffusion, we found that the lower bound on the energetic cost for reaching the target is modified. Specifically, the bound depends on the resetting rate through the MFPT. The energetic cost, in the case where drift is directed toward the target, gradually decreases to zero as the resetting rate decreases. We note that if we consider the drift as an additional external manipulation, then we recover the previous result for regular diffusion, i.e., the mean of the displacements caused by manipulations (resetting and drift) per first passage is $L$.

These findings suggest potential extensions to other particle manipulation protocols. We hypothesize that as long as the cost of dragging the particle toward the target remains linear with distance, the mean sum of displacements per first passage for any manipulation protocol will be L. However, for nonlinear cost functions, we anticipate this cost to be protocol-dependent.
A natural progression of our work involves exploring alternative underlying processes beyond diffusion, such as Lévy flights or run-and-tumble dynamics. These processes, with their blend of deterministic and random characteristics, could potentially be leveraged to reduce the cost per first passage, especially if known a priori and incorporated into the smart resetting protocol.
Finally, we propose a possible connection between our results and the field of finite-time thermodynamics. It would be valuable to investigate how the lower bound on energetic cost achieved by smart resetting relates to the optimal protocol for displacing a colloidal particle between two locations in finite time, as previously calculated within the framework of stochastic thermodynamics \cite{Schmiedl2024}.

\begin{acknowledgement}
Yael Roichman acknowledges support from the Israel Science Foundation (grant No. 385/21). This project received funding from the European Research Council (ERC) under the European Union’s Horizon 2020 research and innovation programme (Grant agreement No. 947731). 
\end{acknowledgement}

\begin{suppinfo}

See Supplementary Information for the details of the theoretical derivations.

\end{suppinfo}
\bibliography{biblio}
\end{document}





\setcounter{equation}{0}
\setcounter{figure}{0}
\setcounter{table}{0}
\setcounter{page}{1}
\setcounter{section}{0}
\makeatletter
\renewcommand{\theequation}{S\arabic{equation}}
\renewcommand{\thefigure}{S\arabic{figure}}
\renewcommand{\thepage}{S\arabic{page}}
\renewcommand{\thesection}{S\Roman{section}} 
\renewcommand{\bibnumfmt}[1]{[S#1]}
\renewcommand{\citenumfont}[1]{S#1}

\begin{center}\Large{Supporting Information for: ``Smart Resetting: An Energy-Efficient Strategy for Stochastic Search Processes''}\end{center}

\author{Ofir Tal-Friedman$^{1}$}

\author{Tommer D. Keidar$^{2,3}$}

\author{Shlomi Reuveni$^{2,3}$}
\email{shlomire@tauex.tau.ac.il}

\author{Yael Roichman$^{1,2}$}
\email{roichman@tauex.tau.ac.il}

\affiliation{\noindent \textit{$^{1}$School of Physics \& Astronomy, Raymond and Beverly Sackler Faculty of Exact Sciences, Tel Aviv University, Tel Aviv 6997801, Israel}}

\affiliation{\noindent \textit{$^{2}$School of Chemistry \& The Center for Physics and Chemistry of Living Systems, Tel Aviv University, Tel Aviv 6997801, Israel}}

\affiliation{\noindent \textit{$^{3}$The Raymond and Beverly Sackler Center for Computational Molecular and Materials Science, Tel Aviv University, Tel Aviv 6997801, Israel}}

\maketitle

\tableofcontents

\newpage 
\section{The Propagator of Diffusion Under Smart Resetting}
The Fokker-Plank equation describing smart resetting with instantaneous returns where the initial position is at the origin and the target is positioned at $x=L$, is given by,
\begin{equation}\label{fokker planck}
\begin{split}
    \frac{\partial p(x,t)}{\partial t}&=D\frac{\partial^2 p(x,t)}{\partial x^2}-rp(x,t)\Theta(-x)+\\
    &+r\delta(x)\int_{-\infty}^0 p(x,t) \, dx=\\
    &=D\frac{\partial^2 p(x,t)}{\partial x^2}-rp(x,t)\Theta(-x)+r\delta(x)Q(t),
\end{split}
 `   \end{equation}
where $D$ is the diffusion coefficient, $r$ is the resetting rate, $\Theta(x)$ is Heaviside function, and $Q(t)\equiv\int_{-\infty}^0 p(x,t) \, dx$. The second term on the right-hand side accounts for probability loss due to resetting, and the third term for probability gain at the origin. The initial condition is that the particle is delta distributed around the origin: $p(x,0)=\delta(x)$. Also, there is an absorbing boundary at $x=L$, thus $p(L,t)=0$.

Laplace transforming Eq. (\ref{fokker planck}) gives
\begin{equation}\label{fokker planck laplace}
    \frac{\partial^2\Tilde{p}(x,s)}{\partial x^2}=\frac{r\Theta(-x)+s}{D}\Tilde{p}(x,s)-\frac{r\Tilde{Q}(s)+1}{D}\delta(x).
\end{equation}
We will solve this equation separately for $x<0$ and $0<x<L$. For $x<0$ we have,
\begin{equation}\label{pde negative}
    \frac{\partial^2\Tilde{p}_-(x,s)}{\partial x^2}=\frac{r+s}{D}\Tilde{p}_-(x,s).
\end{equation}
Taking the general solution and applying the constraint  $p(x\to-\infty,t)=0$, we arrive at,
\begin{equation}\label{pde negative solved}
    \Tilde{p}_-(x,s)=Ae^{\sqrt{\frac{r+s}{D}}x}.
\end{equation}
In the regime $0<x<L$, Eq. (\ref{fokker planck laplace}) simplifies to,
\begin{equation}\label{pde positive}
    \frac{\partial^2\Tilde{p}_+(x,s)}{\partial x^2}=\frac{s}{D}\Tilde{p}_+(x,s).
\end{equation}
The solution for this equation with the boundary condition, $p(L,t)=0$ is given by
\begin{equation}\label{pde positive solved}
    \Tilde{p}_+(x,s)=C[e^{\sqrt{\frac{s}{D}}x}-e^{\sqrt{\frac{s}{D}}(2L-x)}].
\end{equation}
To find the unknown constants we demand continuity of the probability density function at $x=0$. We thus have $p_-(0,t)=p_+(0,t)\Rightarrow \Tilde{p}_-(0,s)=\Tilde{p}_+(0,s)$, which yields
\begin{equation}\label{eqn:appendix A contniuty}
    A=C\left(1-e^{2L\sqrt{\frac{s}{D}}}\right).
\end{equation}
Next, we integrate Eq. (\ref{fokker planck laplace}) over the interval $[-\varepsilon,\varepsilon]$ and take the limit $\varepsilon\to0$. This gives
\begin{equation}\label{integral condition}
     \frac{\partial\Tilde{p}_+(x,s)}{\partial x}\bigg\rvert_{x\to0}-\frac{\partial\Tilde{p}_-(x,s)}{\partial x}\bigg\rvert_{x\to0}=-\frac{r\Tilde{Q}(s)+1}{D}.
\end{equation}
We evaluate the terms on the right-hand side of Eq. (\ref{eqn:appendix A contniuty}) by taking the derivative of Eqs. (\ref{pde negative solved}) and (\ref{pde positive solved}) with respect to $x$, and evaluating at $x=0$. Next, we substitute Eq. (\ref{eqn:appendix A contniuty}) to obtain an equation for the unknown constant $C$. This reads
\begin{equation}
\begin{split}
     &C\sqrt{\frac{s}{D}}\left(1+e^{2L\sqrt{\frac{s}{D}}}\right)- C\sqrt{\frac{r+s}{D}}\left(1-e^{2L\sqrt{\frac{s}{D}}}\right)=\\
     &=-\frac{r\Tilde{Q}(s)+1}{D},
\end{split}
\end{equation}
and solving for $C$ gives
\begin{equation}\label{eqn:appendix A C solution}
    C=\frac{r\Tilde{Q}(s)+1}{\sqrt{D}\left[\sqrt{r+s}\left(1-e^{2L\sqrt{\frac{s}{D}}}\right)-\sqrt{s}\left(1+e^{2L\sqrt{\frac{s}{D}}}\right)\right]}.
\end{equation}
We now observe that
\begin{equation}\label{Q in terms of C}
\begin{split}
    Q(t)&=\int_{-\infty}^0 p(x,t)\, dx \Rightarrow\Tilde{Q}(s)=\int_{-\infty}^0 \Tilde{p}_-(x,s)\, dx=\\
    &=C\left(1-e^{2L\sqrt{\frac{s}{D}}}\right)\sqrt{\frac{D}{r+s}},
\end{split}
\end{equation}
and substituting this result back into Eq. (\ref{eqn:appendix A C solution}) gives
\begin{equation}\label{C solution}
    C=\frac{1}{\sqrt{Ds}\left[\sqrt{\frac{s}{s+r}}\left(1-e^{2L\sqrt{\frac{s}{D}}}\right)-1-e^{2L\sqrt{\frac{s}{D}}}\right]}.
\end{equation}
Finally, we obtain a closed-form expression for the propagator in Laplace space. This reads
\begin{equation}\label{propagator_in_laplace_space}
\begin{split}
    &\Tilde{p}(x,s)=\\
    &=\begin{cases}
        \frac{e^{\sqrt{\frac{s}{D}}(2L-x)}-e^{\sqrt{\frac{s}{D}}x}}{\sqrt{sD}\left[e^{2\sqrt{\frac{s}{D}}L}\left(1+\sqrt{\frac{s}{s+r}}\right)+1-\sqrt{\frac{s}{s+r}}\right]}& \text{if } 0<x<L,\\
        \frac{\left(e^{2\sqrt{\frac{s}{D}}L}-1\right)e^{\sqrt{\frac{s+r}{D}}x}}{\sqrt{sD}\left[e^{2\sqrt{\frac{s}{D}}L}\left(1+\sqrt{\frac{s}{s+r}}\right)+1-\sqrt{\frac{s}{s+r}}\right]} &\text{if } x<0.
    \end{cases}
\end{split}
\end{equation}

\section{First Passage Time Distribution and Mean}
To compute the FPT distribution we will compute the survival probability in Laplace space. The survival probability $\Psi(t)$ is the probability that a particle was not absorbed until time $t$. Its Laplace transform $\Tilde{\Psi}(s)$ can be found by taking the integral of Eq. (\ref{propagator_in_laplace_space}) over all space. This gives
\begin{equation}
    \Tilde{\Psi} (s)=\frac{1}{s+\sqrt{s(s+r)}\coth{\gamma}}+\frac{\cosh{\gamma}-1}{s\left(\cosh{\gamma}+\frac{s}{\sqrt{s(s+r)}}\sinh{\gamma}\right)},
\end{equation}
where $\gamma\equiv\sqrt{\frac{s}{D}}L$. The MFPT can be found by taking the time integral of the survival function
\begin{equation}\label{formula for MFPT}
\begin{split}
    \langle T\rangle &= \int_0^\infty\Psi(t)\,dt=\lim_{s\to0^+} \Tilde{\Psi}(s)=\\
    &=\frac{L^2}{2D}+\frac{L}{\sqrt{Dr}}.
\end{split}
\end{equation}
This result was obtained using a different approach in \cite{catastrophic_resetting}. 

It is instrumental to compare the result in Eq. (\ref{formula for MFPT}) with the MFPT for a diffusing particle under regular stochastic resetting \cite{evans_diffusion_2011}. This reads 
\begin{equation}\label{simple reset MFPT}
    \langle T^{SR}\rangle = \frac{e^{\sqrt{\frac{r}{D}}L}-1}{r}.
\end{equation}
Expanding the above equation to a power series in the resetting rate, $r$, gives
\begin{equation}\label{eqn: appendix B MFPT SR power series}
    \langle T^{SR}\rangle=\frac{L^2}{2D}+\frac{L}{\sqrt{Dr}}+\frac{1}{r}\sum_{n=3}^\infty\frac{1}{n!}\left(L\sqrt{\frac{r}{D}}\right)^n. 
\end{equation}
From here, it is clear that the MFPT in the case of smart resetting is strictly smaller than that of regular resetting. Note, that the results coincide in the limit $r\to0$. In this limit, resetting is only performed rarely. As a result, the chance of finding the particle in the interval $[0, L]$ (given that it survived) is negligible compared to the chance of finding it at $(-\infty, 0)$. There, for all practical purposes, there is almost no difference between regular and smart resetting. 

In the limit $r\to\infty$, the MFPT under regular stochastic resetting diverges. This is because in this limit the density has no time to evolve between resetting events, and it remains concentrated on the origin. For smart resetting --- a completely different behavior is observed. The high resetting rate which is experienced at positions $x<0$ prevents the particle from exploring this part of the line and effectively serves as a reflecting boundary at $x=0$. Any time the particle crosses to the left of the origin --- it is reset back to it in negligible time. Thus, the MFPT of smart resetting in this limit is the same as the one attained for a particle diffusing to an absorbing boundary at $L$, with a reflective boundary at the origin.

\section{Energetic cost for smart resetting}
\label{sec:EnergyCost}
We denote the energy consumption rate due to resetting by $\lambda(t;r)$. The mean of the total energy expended until the target is reached is then given by integrating $\lambda(t;r)$ over time,
\begin{equation}\label{tot energy}
    \langle E_\text{FP}(r)\rangle=\int_0^\infty\lambda(t;r)\, dt.
\end{equation}
Specifically, when the energy consumed in a resetting event is a linear function of the distance traveled, we have
\begin{equation}\label{lambda}
    \lambda(t;r)=-rF_0\int_{-\infty}^0 xp(x,t;r)\, dx,
\end{equation}
where $F_0$ is a general prefactor that, in the setting discussed in the main text, can be interpreted as the force used to displace the particle against fluid friction. Substituting Eq.~(\ref{lambda}) into Eq.~(\ref{tot energy}) and noting its relation to the Laplace transform of the probability distribution, we have
\begin{equation}\label{Tot energy laplacec}
    \langle E_\text{FP}(r)\rangle=-rF_0\int_{-\infty}^0 x\Tilde{p}(x,s\to0;r)\, dx.
\end{equation}
To proceed, we denote the integration factor $A$ that appears in Eq. (\ref{pde negative solved}) as $A(s)$ to highlight its $s$ dependence. Using Eq. (\ref{eqn:appendix A contniuty}) and Eq. (\ref{C solution}) we can write
\begin{equation}\label{eqn: appendix c A_s}
    A(s)=\frac{1-e^{2L\sqrt{\frac{s}{D}}}}{\sqrt{Ds}\left[\sqrt{\frac{s}{s+r}}\left(1-e^{2L\sqrt{\frac{s}{D}}}\right)-1-e^{2L\sqrt{\frac{s}{D}}}\right]}.
\end{equation}
In the limit $s\to0$ Eq. (\ref{eqn: appendix c A_s}) gives $L/D$. By substituting this limit in Eq. (\ref{pde negative solved}), we get that for $x<0$ the Laplace transform of the probability density function at $s=0$ is $\Tilde{p}(x,0)=\frac{L}{D}e^{\sqrt{\frac{r}{D}}x}$. So by using this result and Eq. (\ref{Tot energy laplacec}) we can see that the mean energy consumption is given by
\begin{equation}
\begin{split}
    \langle E_\text{FP}(r)\rangle&=-\frac{rF_0L}{D}\int_{-\infty}^0xe^{\sqrt{\frac{r}{D}}x}\, dx=\\
    &=F_0L\Gamma(2)=F_0L,
\end{split}
\end{equation}
where $\Gamma(x)$ is the gamma function. Surprisingly, the mean energy consumption is constant and equal to the work needed to simply drag the particle from the origin to the target using the same force.

In a more general setting, where the energy for dragging the particle a distance $\ell_0$ is $\kappa \ell_0^\alpha$ the mean energy required in the process is given by,

\begin{equation}\label{energy gen power law}
\begin{split}
    \langle E_\text{FP}(r)\rangle&=\frac{r\kappa L}{D}\int_{-\infty}^0 (-x)^\alpha e^{\sqrt{\frac{r}{D}}x} \,dx = \\
    &=\frac{r\kappa L}{D}\left(\frac{D}{r}\right)^{\frac{\alpha+1}{2}}\Gamma(\alpha+1)=\\
    &=\kappa L\Gamma(\alpha+1)\left(\frac{D}{r}\right)^{\frac{\alpha-1}{2}}  
\end{split}
\end{equation}

Comparing the result above to Eq. (48) in \cite{Sunil_2023} we can see that for any cost which is a power law in the dragging distance, the cost for stochastic resetting and smart resetting are asymptotically equal in the limit $r\to0$. This is to be expected from the equivalence between the two resetting strategies in this limit. What is somewhat surprising is the fact that the cost of smart resetting at \emph{any} resetting rate, $r$, has the same functional form as the cost of regular resetting at low resetting rates. In particular, the finite (asymptotic) cost obtained for regular stochastic resetting in the limit $r\to0$ where the cost is linear in the dragging distance, holds for all resetting rates in the case of smart resetting. 

\section{Mean Drag Distance Toward The Target in Regular Resetting}

In a regular resetting protocol, the particle is sometimes dragged toward the target and sometimes away from it. If the particle is in the interval $(0,L)$ it will be dragged away from the target, and if it is in the semi-infinite line $(-\infty,0)$ it will be dragged toward it. The average of the total distance a particle is dragged can be written in the following way \cite{tal-friedman_experimental_2020, Sunil_2023,pal2020search}
\begin{equation}\label{total distance as linear combination}
    \langle \ell\rangle = \langle \ell_t \rangle+\langle \ell_a\rangle = 2\sqrt{\frac{D}{r}}\sinh{\left(\sqrt{\frac{r}{D}}L\right)}-L
\end{equation}
where $\langle \ell_t \rangle$, $\langle \ell_a\rangle $ are the distance dragged toward and away from the target respectively. We would like to calculate the contribution of $\langle \ell_t \rangle$ to Eq. (\ref{total distance as linear combination}). We will do so using the following description of the cost of a resetting process. The particle starts diffusing, and at a random time its location is measured. If the particle has not reached the target yet, it is reset, a cost $c(x)$ is paid where $x$ is its current location, and a new independent and identically distributed measuring time is drawn. On the other hand, if the particle has already reached the target before its measurement, no cost is paid, and no more measurements are done. 

To compute the total cost of such a process, one needs to compute the mean cost of a single measurement event where the cost is 0 if the particle already reached the target and $c(x)$ otherwise. The mean cost of a single measurement can then be multiplied by the mean number of measurements to obtain the mean total cost.

The mean cost of a single measurement event at time $t$ can be calculated using the total expectation theorem
\begin{equation}\label{eq: cost of single drag}
    \langle c(t)\rangle = \Psi(t)\int_X c(x)\frac{G(x,t)}{\Psi(t)}\, dx=\int_X c(x)G(x,t)\, dx, 
\end{equation}
where $X$ is the total available space, $\Psi(t)$ is the survival probability of the process, and $G(x,t)$ is the propagator of the underlying process in the presence of an absorbing boundary. Namely, the probability of finding a particle at location $x$ at time $t$. To find the mean cost of a measurement event, eq. (\ref{eq: cost of single drag}) should be averaged over the probability density function of the measuring times. In our case, the measuring times are exponentially distributed with a rate $r$ resulting in
\begin{equation}\label{eq: mean cost per single}
    \langle c\rangle = r\int_0^\infty e^{-rt}\langle c(t)\rangle\, dt = r\int_X c(x)\Tilde{G}(x,r)\, dx,
\end{equation}
where $\Tilde{G}(x,r)$ is the Laplace transform of $G(x,t)$ evaluated at $r$. The number of measurements until the last one is geometrically distributed, with a success probability $Pr(T<R)$, i.e., that first-passage will happen before the measurement. Therefore, the mean number of measurements is $1/Pr(T<R)$. In our case of exponentially distributed measuring times
\begin{equation}\label{eq: success probability}
    Pr(T<R)=\int_0^\infty f_T(t)e^{-rt}\, dt=\Tilde{T}(r),
\end{equation}
where $f_T(t)$ is the FPT probability density function, and $\Tilde{T}(r)$ is its Laplace transform evaluated at $r$.

Coming back to our goal, we want to find the mean distance dragged toward the target until a first-passage when resetting a diffusive particle. Therefore we need to use a cost function $c_t(x)$ such that its value will be the distance to the origin when the origin is between the particle to the target, and zero otherwise. For the case dealt with in this paper, where the target is located at $L>0$, this function is

\begin{equation}\label{cost function for dragging toward}
    c_t(x)=\begin{cases}
        0 &\text{if $0<x<L$},\\
        -x&\text{if $x<0$}.
    \end{cases}
\end{equation}
By using the above cost function in eq. (\ref{eq: mean cost per single}), and divide by $\Tilde{T}(r)$ according to eq. (\ref{eq: success probability}), we arrive at the following expression for the mean distance dragged toward the target in stochastic resetting

\begin{equation}\label{renewal mean cost toward laplace}
    \langle \ell_t\rangle = \frac{-r\int_{-\infty}^0x\Tilde{G}(x,r)\, dx}{\Tilde{T}(r)}.
\end{equation}
For free diffusion with an absorbing boundary at $L$, and for $x<0$ we have $\Tilde{G}(x,r)=\frac{1-e^{-\sqrt{\frac{4r}{D}}L}}{\sqrt{4Dr}}e^{\sqrt{\frac{r}{D}}x}$ and $\Tilde{T}(r)=e^{-\sqrt{\frac{r}{D}}L}$. Plugging those functions to eq. (\ref{renewal mean cost toward laplace}) results in the following expression
\begin{equation}\label{mean distance toward}
    \langle \ell_t\rangle = \sqrt{\frac{D}{r}}\sinh{\left(\sqrt{\frac{r}{D}}L\right)}.
\end{equation}
We see that the total dragging distance toward the target in regular stochastic resetting is bounded from below by the total dragging distance in smart resetting. The absence of feedback from the system in regular resetting increases the mean total drag distance in two ways: The particle is dragged away from the target which increases the dragging distance while increasing the MFPT, and the dragging of particles away from the target increases their density in the infinite half line behind the initial position. In turn, the increasing density in the negative half-line increases the dragging distance toward the target.

\section{MFPT of Drift-Diffusion with Smart Resetting}
\label{Sec:MGPT_DD}

We will use the approach presented in \cite{catastrophic_resetting} to compute the MFPT for drift-diffusion under smart resetting. The FPT statistics of the problem presented here, where a particle starts from the origin and there is an absorbing boundary at $L>0$, will be the same as the one obtained for a particle starting at $L>0$ and drift toward the origin where the drift sign is flipped. In \cite{catastrophic_resetting} it was shown, using a backward Kolmogorov approach, that the MFPT of those kinds of problems with smart resetting can be obtained by solving the following ODE, and evaluating it at $x=L$
\begin{equation}\label{eq: MFPT ODE}
    \frac{d^2\tau}{dx^2}-\frac{v
    }{D}\frac{d\tau}{dx}+\frac{r}{D}\Theta(x-L)\left[\tau(L)-\tau(x)\right]=-\frac{1}{D},
\end{equation}
where $\Theta(y)$ is Heaviside step function. We kept our original notation where positive $v$ indicates drift toward the target. The equation needed to be solved in the interval $[0,\infty)$ with the boundary conditions that $\tau(0)=0$, and $\lim_{x\to\infty}\tau(x)$ is finite. Moreover, we will require that $\tau$ and $\frac{d\tau}{dx}$ are continuous functions. The solution for $x\leq L$ is
\begin{equation}
    \tau_{x\leq L}(x)=A\left(1-e^{\frac{vx}{D}}\right)+\frac{x}{v},
\end{equation}
where $A$ is an integration constant that will be found later. For $x>L$, we get the following solution
\begin{equation}
    \tau_{x>L}(x)=Be^{\alpha_-x}+Ce^{\alpha_+x}+\frac{1}{r}+\tau(L),
\end{equation}
where $\alpha_\pm\equiv\frac{1}{2D}(v\pm\sqrt{v^2+4rD})$. Because $\lim_{x\to\infty}\tau(x)$ is finite, we must set $C=0$. By requiring self consistency on the value of $\tau(L)$, we get
\begin{equation}
    \tau_{x>L}(x)=\frac{1}{r}\left(1-e^{\alpha_-(x-L)}\right)+\tau(L).
\end{equation}
Using the continuity of $\frac{d\tau}{dx}$ at $x=L$, we get that
\begin{equation}
    A=e^{-2Pe}\left[\frac{D}{v^2}+\frac{1-\frac{|v|}{v}\sqrt{1+\frac{4rD}{v^2}}}{2r}\right],
\end{equation}
where $Pe\equiv\frac{Lv}{2D}$ is the Péclet number. 
Therefore,
\begin{equation}
\begin{split}
    \tau(L)&=\langle T\rangle\\
    &=\left[\frac{D}{v^2}+\frac{1-\frac{|v|}{v}\sqrt{1+\frac{4rD}{v^2}}}{2r}\right]\left(e^{-2Pe}-1\right)+\frac{L}{v}.
\end{split}
\end{equation}

\section{Corroboration of Eq. (14)}
\label{Sec:plot_drift}
Using Langevin simulations and the MFPT for drift-diffusion under regular and smart resetting, we corroborate Eq. (14) from the main text,
 \begin{equation} \label{eqn: w_t-w_a=L drift appendix}
    \langle \ell_t\rangle-\langle \ell_a\rangle=L-v\langle T \rangle.
\end{equation}

\begin{figure}[h]
	\centering
	\includegraphics[trim={4.55cm 0 3.5cm 0},clip,width=0.5\textwidth]{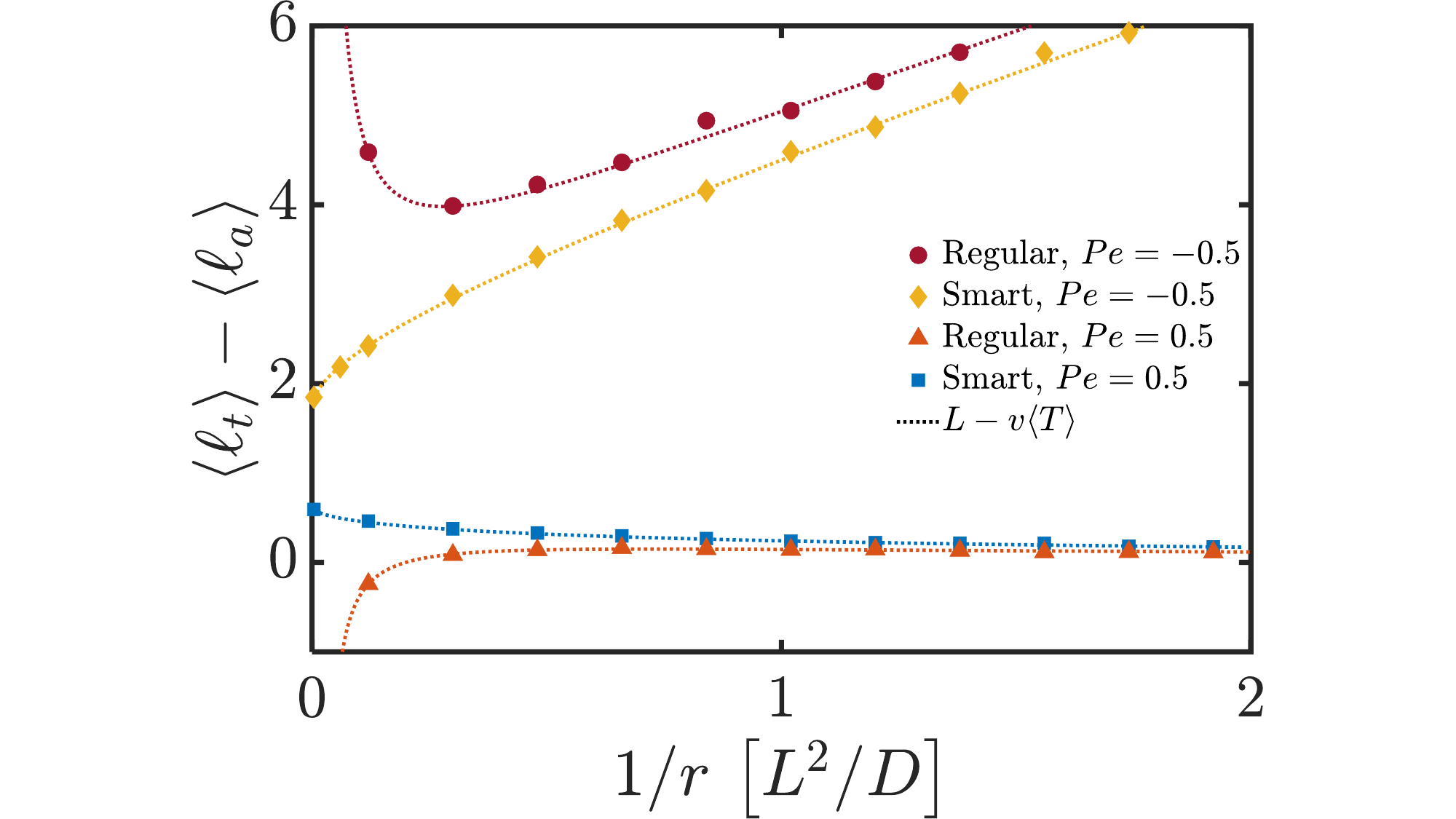}
	\caption{Eq. \eqref{eqn: w_t-w_a=L drift appendix} plotted for regular and smart resetting, with positive and negative Péclet numbers. For all cases, there is a good agreement between $\langle \ell_t\rangle-\langle \ell_a\rangle$ and $L-v\langle T \rangle$.}
 \label{fig:appendix_fig}
 \end{figure}
 
 \begin{figure}[h]
	\centering
	\includegraphics[trim={3.9cm 0 3.2cm 0},clip,width=0.5\textwidth]{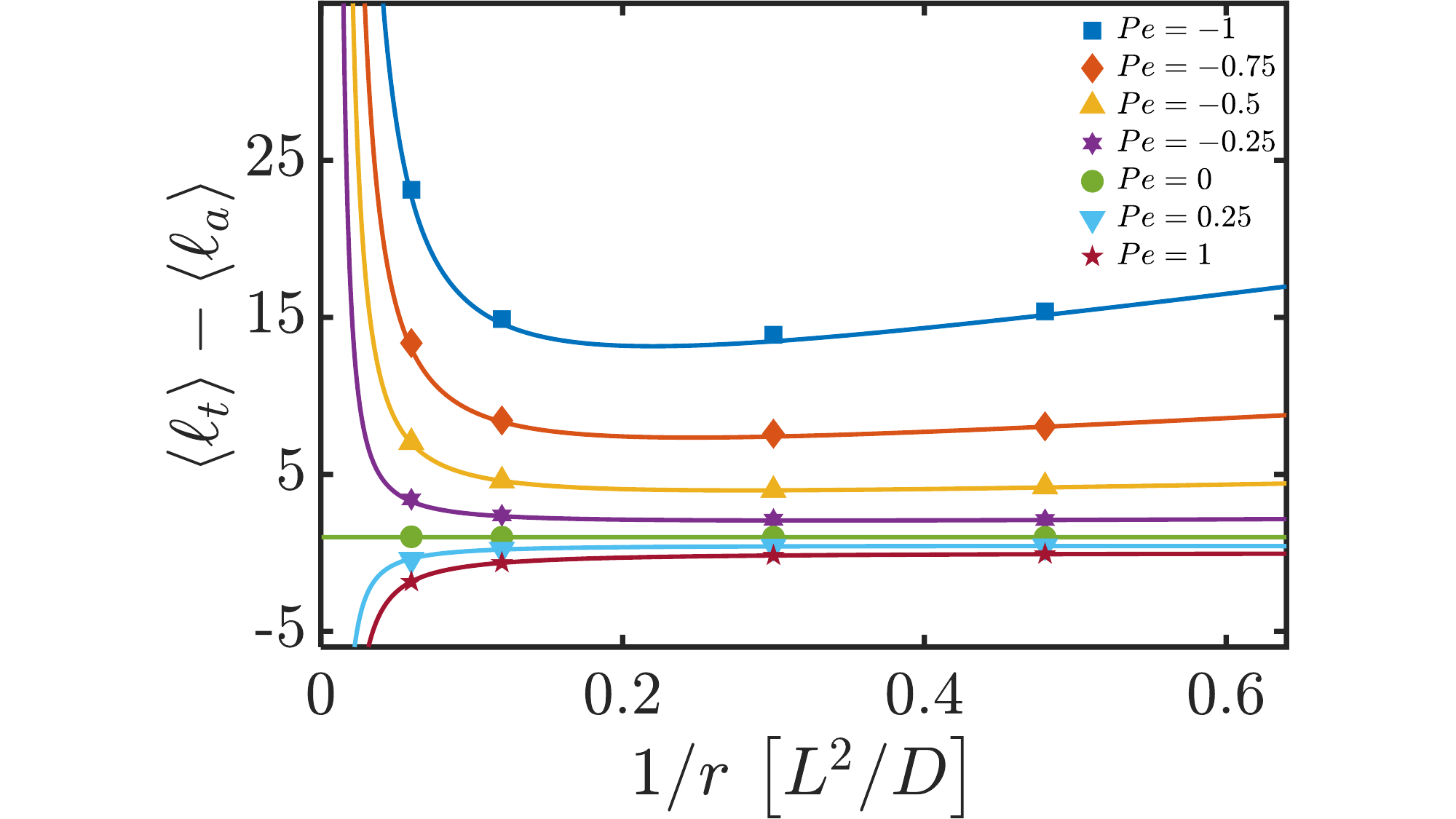}
	\caption{Eq. \eqref{eqn: w_t-w_a=L drift appendix} plotted for regular resetting, with various Péclet numbers. For all cases, there is a good agreement between $\langle \ell_t\rangle-\langle \ell_a\rangle$ and $L-v\langle T \rangle$.}
 \label{fig:appendix_only_regular_fig}
 \end{figure}

\bibliography{biblio}